\documentclass[10pt,letterpaper]{article}
\usepackage{opex3}
\usepackage{cite}

\usepackage{ae} 

\begin{document}

\title{Frequency dependence of mode coupling gain in Yb doped fiber amplifiers due to stimulated thermal Rayleigh scattering}

\author{Arlee V. Smith$^*$ and Jesse J. Smith}

\address{AS-Photonics, LLC, 8500 Menaul Blvd. NE, Suite B335, Albuquerque, NM USA 87112}

\email{$^*$arlee.smith@as-photonics.com}

\begin{abstract}
Using a numerical model we study the frequency dependence of mode coupling gain due to stimulated thermal Rayleigh scattering in step index, Yb doped, fiber amplifiers. The frequency at the gain peak is shown to vary with core size, doping size, population saturation, thermal lensing, fiber coiling, direction of pumping, photodarkening, and pump noise spectra. The predicted frequencies are compared with measured values whenever possible.
\end{abstract}


\section{Introduction}

In earlier papers we described the physical process that causes modal instability\cite{smithsmith2011} in high power fiber amplifiers. A key aspect of this stimulated thermal Rayleigh scattering (STRS) process is a frequency offset between the two coupled modes, usually LP$_{01}$ and LP$_{11}$. We have incorporated the physics of this STRS process in a detailed, highly numerical model\cite{smithsmith2013} that can compute mode coupling gain as a function of the frequency offset. A typical gain curve is shown in Fig. 1. In \cite{smithsmith2011} we argued that the frequency of peak gain could be roughly estimated as the inverse of the thermal diffusion time across the core radius. This implies the peak gain for typical large mode area fibers should lie in the frequency range 200-5000 Hz.
\begin{figure}[hpb]
\centering
\includegraphics[width=4.5in]{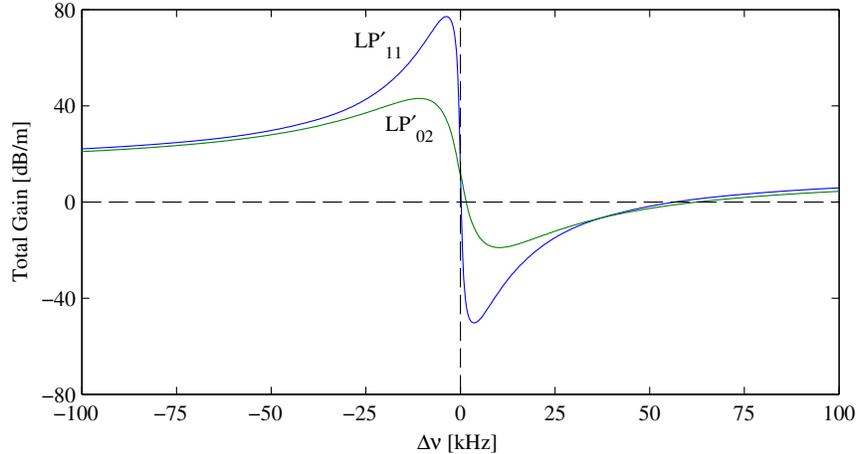}
\caption{\label{fig:gaincurve}Mode coupling gain plus lasing gain versus frequency offset for a large mode area fiber amplifier. The coupling is from LP$_{01}$ to either LP$_{11}$ or LP$_{02}$. The peak value of gain and its position vary with amplifier parameters.}
\end{figure}

Semi analytic versions of our model of STRS have been presented by Hansen{\it et al.}\cite{hansen,hansen2}. 
Hansen {\it et al.} also computed gain versus frequency offset, but with the assumptions that the transverse heat profile is proportional to the transverse irradiance profile and that the mode profiles are constant along the fiber. These simplifying assumptions permit the development of a compact model, but such models miss some of the subtleties of real amplifiers. For instance, in real amplifiers the shape of the mode coupling gain $g(z)$ is altered by any change in the transverse heat deposition profile along the fiber, and depends not only on the core refractive index step and the doping profile, as found by Hansen {\it et al.}, 
 but also on the bend radius, linear absorption of the signal due to photodarkening or other processes, thermal lensing, and with total thermal load if cooling is asymmetric. However, perhaps the most important difference between the simplified models and our more detailed model is related to changes in the transverse profile of the population inversion, or transverse hole burning, which dramatically alters the shape of the oscillating heat source responsible for STRS as the balance between pump and signal powers changes. This is most noticeable in co-pumped amplifiers where the pump power falls while the signal power rises as they propagate down the fiber. However, hole burning is also important in counter-pumped amplifiers, even though the ratio of pump to signal powers is more nearly constant along the fiber. Population depletion affects the magnitude of the mode coupling gain as well as its frequency profile, but its influence on the magnitude of gain will be the topic of other papers.

In this report we present modeling results for amplifiers operated at the gain peak and near the instability threshold. We expect our steady-periodic model to work well near threshold although it may be less useful for operation well above threshold. Here we are primarily concerned with how the frequency that gives maximum gain near the instability threshold ($F_M$) changes with fiber design and operating conditions. Besides being an important test of our model, this information will be useful in future modeling since it reduces the scope of the two dimensional search required to find $F_M$ at the threshold. We also anticipate that measured values of $F_M$ will be useful in diagnosing causes of anomalously performance such as low instability thresholds. As we will show, different problems can have different influences on the frequency. Even so, it is unlikely that in practice frequency information alone will uniquely identify such problems. 

\section{Frequency vs mode field diameter (or effective area) for heat profile proportional to irradiance profile}

In this section we show how $F_M$ varies with core size, using the simplifying assumption that the heat profile mimics the irradiance profile. This is the same approximation used by Hansen {\it et al.}\cite{hansen}. 
It gives a preliminary estimate of $F_M$ that we will improve on in the next section. For this preliminary computation we use a small linear signal absorption inside the doped region, with no pump absorption. This produces a heat profile proportional to the irradiance profile in the doped region. All the power lost to linear absorption is assumed to appear as heat that causes mode coupling via STRS. We use a step index fiber with numerical aperture ($NA$) of 0.054 and a 976 nm pump. Our peak frequencies for the 20, 40, and 80 $\mu$m diameters agree closely with those of Hansen {\it et al.}\cite{hansen} even though we use a 1040 nm signal while Hansen uses 1030 nm. This is expected because $F_M$ depends only on the shapes of the two modes and they do not change significantly over this wavelength range. Our results are summarized in Fig. 2.
\begin{figure}[hpb]
\centering
\includegraphics[width=4.5in]{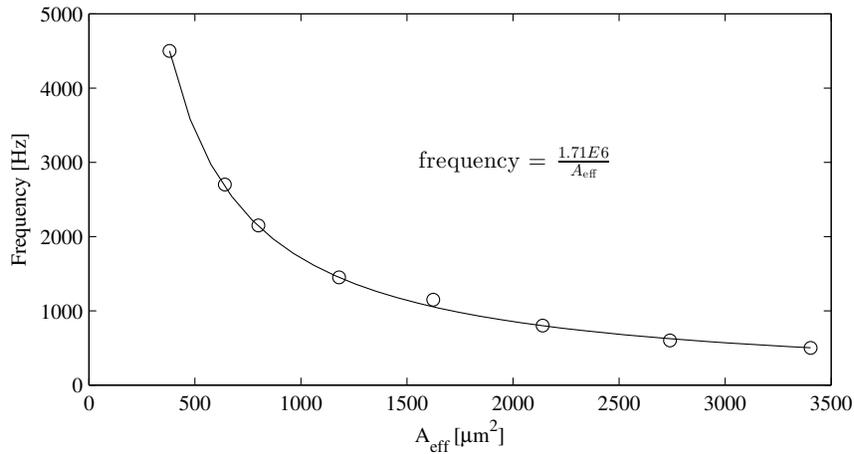}
\caption{\label{fig.freqvaeff}Frequency of peak gain ($F_M$) {\it vs} $A_{\rm eff}$ of mode LP$_{01}$ for a constant numerical aperture of 0.054 and a 1040 nm signal. The heat profile matches the irradiance profile over the doped region in this case. The circles are found using our numerical model for core diameters of \{25 35 40 50 60 70 80 90\} $\mu$m with corresponding effective areas of \{380 645 805 1180 1630 2155 2755 3420\}, while the solid line is given by the best fit inset equation.}
\end{figure}

The frequencies are well approximated by the equation
\begin{equation}\label{eq.171}
F_M=1.71\times 10^6/A_{\rm eff}.
\end{equation}
where $A_{\rm eff}$ is the effective area of the LP$_{01}$ mode in square microns and the signal wavelength is 1040 nm. This result can be compared with a frequency estimate based on the thermal diffusion time for a heated cylinder of radius $r_{\rm eff}$
\begin{equation}
F = \frac{K}{r_{\rm eff}^2C\rho}
\end{equation}
which for silica thermal parameters ($K=1.38$ W/m-K, $C=703$ J/kg-K, $\rho=2201$ kg/m$^3$) gives
\begin{equation}\label{eq.89}
F = 8.9\times 10^5/r_{\rm eff}^2.
\end{equation}
The two frequencies computed using Eqs. (\ref{eq.171}) and (\ref{eq.89}) are equal if
\begin{equation}
A_{\rm eff}=1.9 r_{\rm eff}^2.
\end{equation}
Defining the mode field radius $r_{\rm mf}$ by
\begin{equation}
A_{\rm eff}=\pi r_{\rm mf}^2
\end{equation}
means
\begin{equation}
r_{\rm eff}=1.28r_{\rm mf},
\end{equation}
a reasonable result. 

In the computations above we fixed the numerical aperture at $NA=0.054$. Changing the $NA$ changes the shape of the modes and thus the frequency. Hansen {\it et al.}\cite{hansen} showed the frequency increases as the $V$ parameter, or equivalently, the $NA$ increases. An increase in $NA$ tends to compress the modes so it is no surprise $F_M$ increases slightly.

\section{Realistic heat profiles}

\begin{figure}[hpb]
\centering
\includegraphics[width=4.5in]{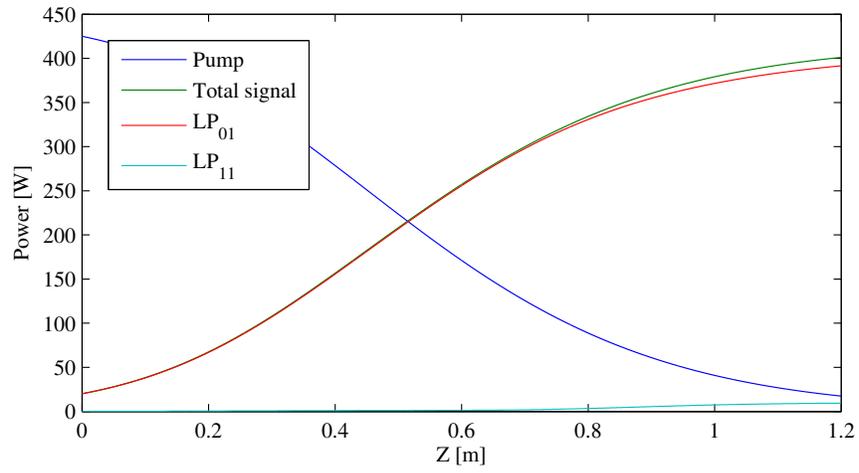}
\caption{\label{fig.copowers}Signal and pump power distributions along typical co-pumped amplifier.}
\end{figure}
\begin{figure}[hpb]
\centering
\includegraphics[width=4.5in]{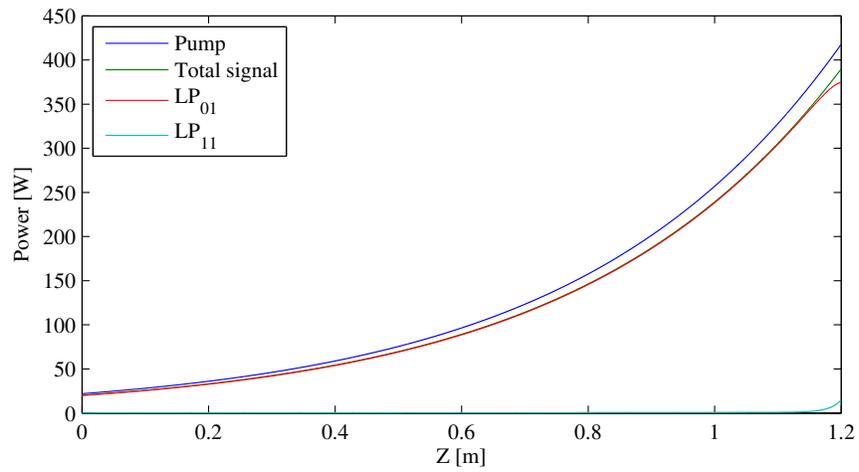}
\caption{\label{fig.counterpowers}Signal and pump power distributions along typical counter-pumped amplifier.}
\end{figure}

\begin{figure}[hpb]
\centering
\includegraphics[width=4.5in]{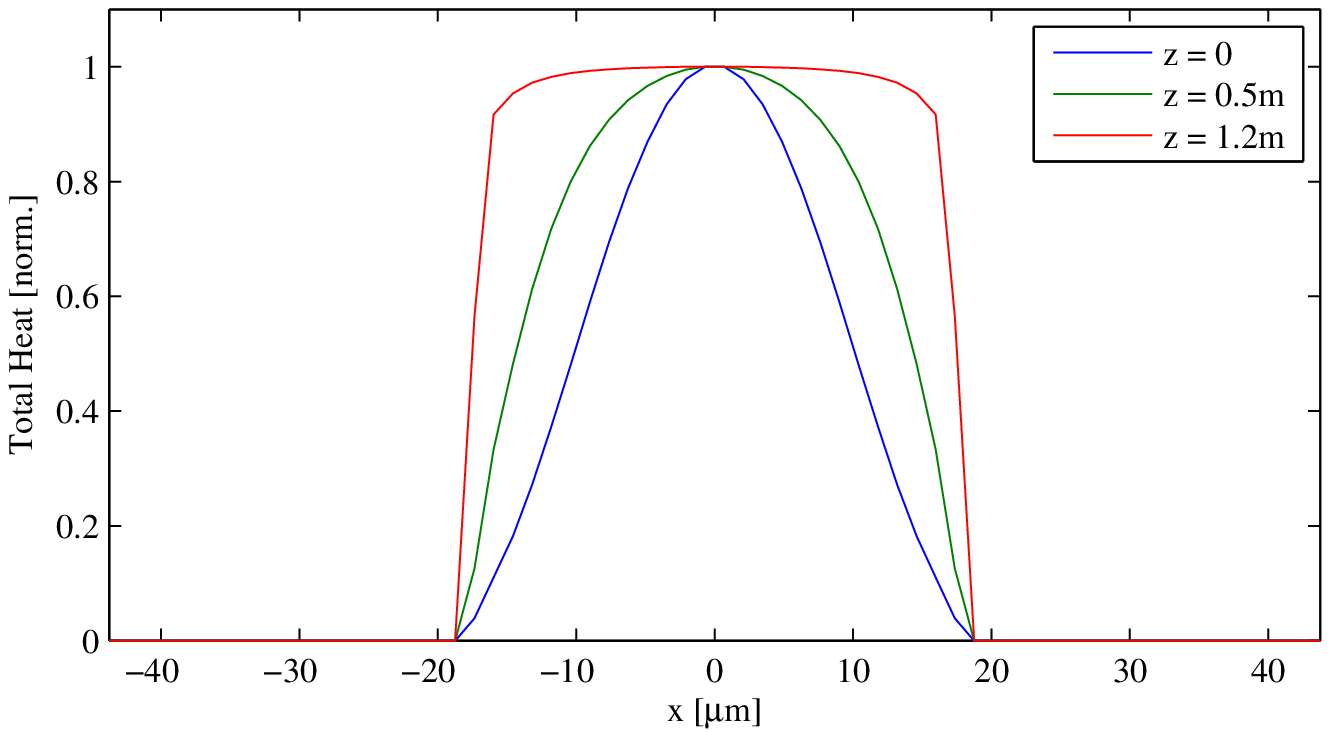}
\caption{\label{fig.totalheat}Transverse cuts through the normalized profiles of the time averaged heat at three $z$ locations along the 1.2 m long, co-pumped fiber. The fiber has $d_{\rm core}=d_{\rm dope}=35$ $\mu$m and is designed for high conversion efficiency with $N_{\rm Yb}=3\times 10^{25}$ m$^{-3}$ and $d_{\rm clad}=140$ $\mu$m. It is co-pumped with 425 W, seeded with 20 W of 1040 nm signal in LP$_{01}$ and 0.2 W in LP$_{11}$. The powers evolve as shown in Fig. \ref{fig.copowers}}
\end{figure}

\begin{figure}[hpb]
\centering
\includegraphics[width=4.5in]{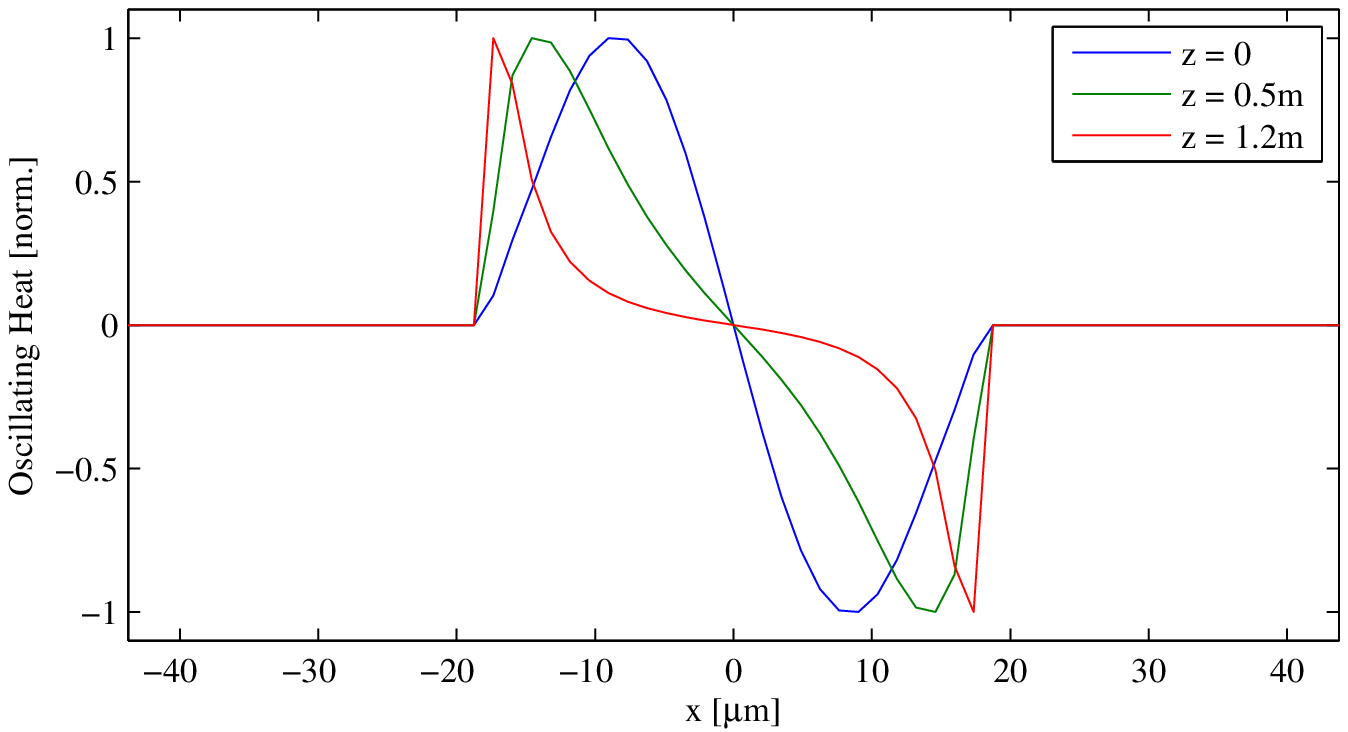}
\caption{\label{fig.oscheat}Transverse cuts through the normalized profiles of the anti symmetric, oscillatory portion of the heat at three locations along the co-pumped fiber. The amplifier is otherwise the same as in Fig. \ref{fig.totalheat}.}
\end{figure}

In real amplifiers the heat profile does not match the product $E_{01}E_{11}$ as assumed above because the upper state population is depleted by varying amounts across the core. The degree of depletion varies with the ratio of signal to pump irradiances, so it also varies along the length of the amplifier. Figures \ref{fig.copowers} and \ref{fig.counterpowers} show how the signal and pump powers vary along the fiber for high efficiency co- and counter-pumped amplifiers. For the co-pumped amplifier the pump and signal powers are approximately equal midway along the fiber. For the counter-pumped amplifier they are approximately equal along most of the fiber. Near the input of the co-pumped fiber the pump is much stronger than the signal so there is little population depletion, or hole burning, and the approximation of the previous section is nearly met, and $F_M$ is given by Eq. (\ref{eq.171}). The time-averaged heat profile at this location is shown in blue in Fig. \ref{fig.totalheat}, while its oscillatory component is shown in blue in Fig. \ref{fig.oscheat}. The corresponding heat profiles midway along the fiber are shown as green curves in the two plots, and the profiles near the output are shown as red curves. For the same amplifier with counter-pumping the profiles are nearly constant along the fiber and are approximately the same as those near the midpoint of the co-pumped fiber.

Clearly, transverse hole burning strongly alters the profile of the anti symmetric, oscillatory part of the heat deposition. The assumption of the previous section which ignored hole burning as unimportant is highly dubious for real amplifiers, whether co- or counter-pumped. The outward displacement of the oscillatory heating means the asymmetry persists longer before it decays away by thermal diffusion, and this means $F_M$ is reduced as hole burning strengthens. For the co-pumped fiber most of the mode coupling gain occurs near the mid point, so the heat distribution at that location largely determines $F_M$. The value of $F_M$ for the counter-pumped fiber is approximately the same because the heat profile over the full fiber length is nearly the same as at the midpoint of the co-pumped fiber.

\begin{figure}[hpb]
\centering
\includegraphics[width=4.5in]{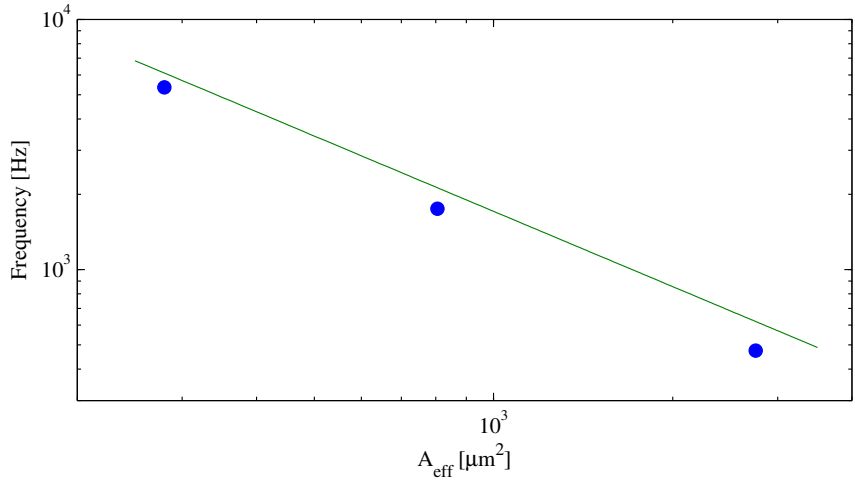}
\caption{\label{fig.realfreqs}Comparison of $F_M$ computed using the simplified (solid line) and full models (circles) for core diameters of 20, 40, and 80 $\mu$m. All three are 1.2 m long, with $NA=0.054$, co-pumped at 976 nm, with $\lambda_s=1040$ nm, and operating near the mode instability threshold.}
\end{figure}

Full model runs using computed population inversions are used to predict realistic frequencies for the 20, 40, and 80 $\mu$m diameter, 1.2 m long, co-pumped amplifiers of Fig. \ref{fig.freqvaeff}. The pump cladding diameter is scaled with the core diameter to maintain similar pump to signal ratios and similar conversion efficiency. The results are shown as the symbols in Fig. \ref{fig.realfreqs}. For the (20, 40, 80) $\mu$m diameter cores the simple model gives (6130, 2125, 620) Hz while the full model gives (5350, 1750, 475) Hz, making the ratios (0.87, 0.82, 0.77). 

\section{Dependence on doping diameter}

\begin{figure}[hpb]
\centering
\includegraphics[width=4.5in]{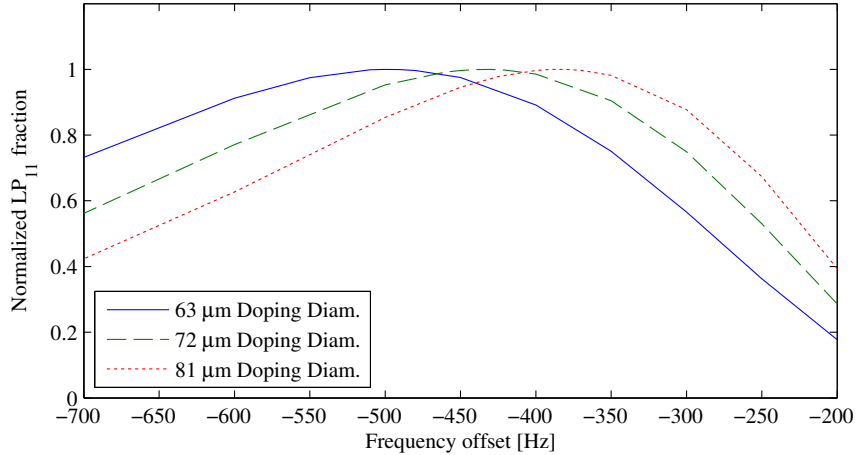}
\caption{\label{fig.doping}Mode LP$_{11}$ fraction at fiber output versus frequency offset for varying doping diameters in co-pumped fiber. The core diameter is fixed at 81 microns and the doping diameters are 81, 72, and 63 $\mu$m; $L=1.2$ m; $d_{\rm clad}=255$ $\mu$m; $N_{\rm Yb}=3\times 10^{25}$ m$^{-3}$. In each case the pump power is adjusted to the mode instability threshold.}
\end{figure}

It can be anticipated that if the Yb$^{3+}$ dopant is confined to a smaller diameter than the core index step, the influence of hole burning on $F_M$ will be less than for a fully doped core because heating will be confined to the doped zone and cannot broaden as much. Keeping other fiber properties constant and varying only the doping diameter, Hansen {\it et al.}\cite{hansen} calculated the frequency for three doping diameters with a fixed refractive index profile. They showed $F_M$ increased with decreasing doping diameter. However, they used heat profiles proportional to the irradiance profiles. We have recomputed the frequency response using our numerical model for one fiber design, varying the doping diameter and adjusting the pump power as necessary to achieve threshold operation. Figure \ref{fig.doping} shows our results.
In all three curves the power is near the mode instability threshold so the fraction of signal power in LP$_{11}$ is approximately 0.3\% at the output for the frequency of highest gain. The fiber is the step index equivalent of LPF45 ($d_{\rm core}=81$ $\mu$m, $d_{\rm clad}=255$ $\mu$m, $L=1.2$ m, $n_{\rm core}=1.45015$, $n_{\rm clad}=1.45$, $NA=0.0209$, and $V=10.2$) with signal input in LP$_{01}$=10 W, signal input in LP$_{11}=10^{-7}$ W. The gain peaks corresponding to doping diameters (63, 72, 81) $\mu$m lie at frequencies (500, 430, 380) Hz. There is probably no deep significance to it, but the product of $F_M$ and $d_{\rm dope}$ is nearly constant for these three cases. 

\section{Dependence on V parameter}

A lower value of the numerical aperture would allow the modes to expand slightly into the cladding, and this should reduce $F_M$. This effect has been verified by Hansen {\it et al.}\cite{hansen}. 
under the assumption of no hole burning.



\section{Dependence on thermal lensing}

\begin{figure}[hpb]
\centering
\includegraphics[width=4.5in]{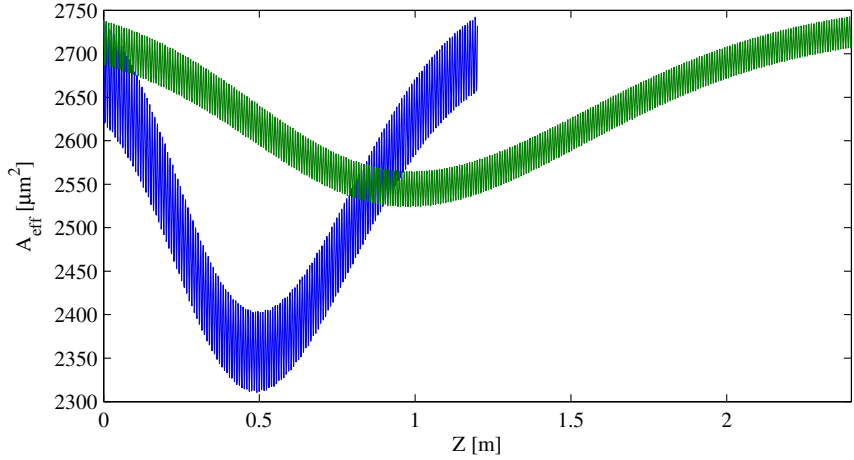}
\caption{\label{fig.lensing}$A_{\rm eff}$ for an amplifier with $d_{\rm core}=d_{\rm dope}=80$ $\mu$m. Other parameters: $NA=0.054$, $d_{\rm clad}=376$ $\mu$m, LP$_{01}$ seed power is 10 W, LP$_{11}$ seed power is $10^{-16}$ W. The rapid oscillations are caused by launch of seed light with a profile equal to the low power mode instead of a self consistent thermally lensed mode.}
\end{figure}

When the signal light is largely confined to the fundamental mode, as it is below and near threshold, the temperature peaks at the center of the core, and this causes thermal lensing, or self focusing of the signal light. The result is a constriction of the mode profiles which might be expected to cause an increase in $F_M$. The thermal lensing effect is stronger in larger diameter fibers. For example, it can be quite significant in an 80 $\mu$m diameter fiber operating at several hundred watts, but it is much less noticeable for a 20 $\mu$m diameter fiber. In Fig. \ref{fig.lensing} we show the effective area of the fundamental mode versus position along the fiber for two cases, a short and a long fiber with identical 80 $\mu$m diameter cores and equal pump cladding sizes. The length of one is 1.2 m, the other is 2.4 m. The core is doped over its full diameter, but at a density of $3\times 10^{25}$ m$^{-3}$ for the short fiber and half that for the long fiber. This makes the hole burning similar for the two, but the heat per length is halved in the longer fiber, reducing its thermal lensing. As the figure shows, lensing reduces $A_{\rm eff}$ near the mid point by 13\% for the shorter fiber and 7\% for the longer fiber. 

\begin{figure}[hpb]
\centering
\includegraphics[width=4.5in]{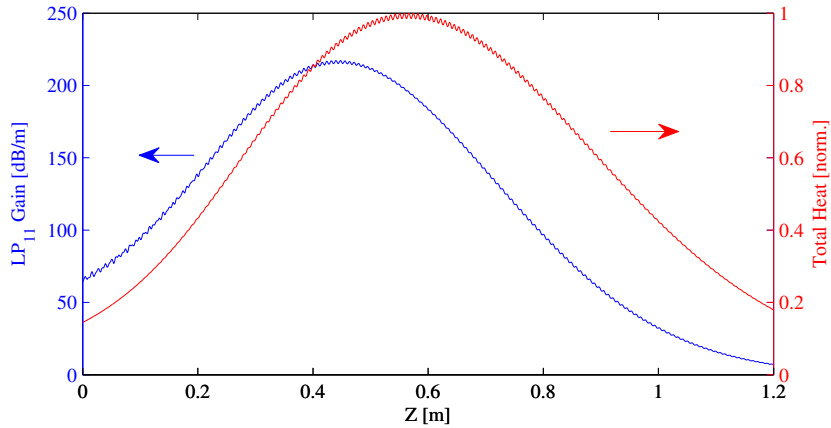}
\caption{\label{fig.gainheat}Blue curve is gain of LP$_{11}$, including laser gain and mode coupling gain versus $z$ for the same fiber as in Fig. \ref{fig.lensing}. $A_{\rm eff}$ for an amplifier with $d_{\rm core}=d_{\rm dope}=80$ $\mu$m. Red curve is the total heat deposited at location $z$ in normalized units.}
\end{figure}

If $F_M$ scaled in proportion of $1/A_{\rm eff}$ this would cause a frequency shift of approximately 40 Hz between the two fibers. In fact, the modeled frequency shift is only 20 Hz, with the longer fiber having a lower frequency, as expected. The heat and LP$_{11}$ gain profiles of the 1.2 m fiber are shown in Fig. \ref{fig.gainheat}. The shift reduction is partly due to the fact that the gain peaks earlier in the fiber than the thermal lensing, and partly due to the fact that the gain is distributed over sections of fiber that have less lensing than the mid point. The result is that the effective frequency reduction is less than guessed based on the maximum reduction in $A_{\rm eff}$. Additionally, the smaller beam in the more strongly lensed short fiber leads to hole burning, and this also reduces $F_M$ by small amount, bringing its shift closer to that of the long fiber. The lesson is that the influence of thermal lensing on $F_M$ is less than might be guessed based only on the reduction in $A_{\rm eff}$.

\section{Dependence on coiling radius}

\begin{figure}[hpb]
\centering
\includegraphics[width=4.5in]{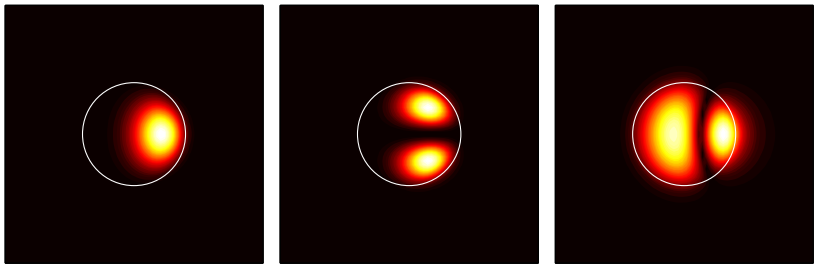}
\caption{\label{fig.coilmode}Irradiance profiles of LP$_{01}$ (left image) and the two orientations of LP$_{11}$ in a step index fiber with numerical aperture 0.054, core diameter of 40 $\mu$m, coiled to a 100 mm radius. LP$_{11o}$ (center image) is the mode with lobes out of the bend plane; LP$_{11i}$ (right image) is the mode with lobes in the bend plane.}
\end{figure}

\begin{figure}[hpb]
\centering
\includegraphics[width=4.5in]{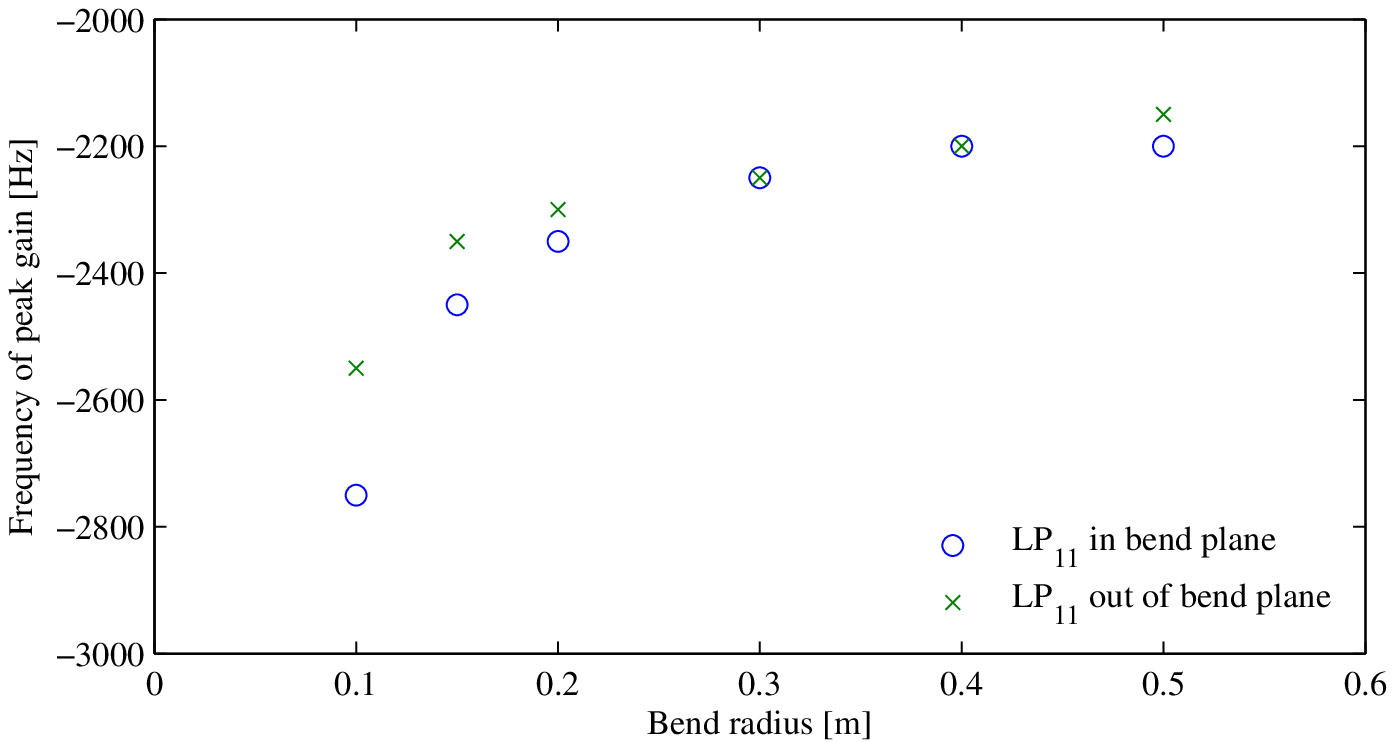}
\caption{\label{fig.coilfreq}Frequency of the gain peak for different coiling radii for coupling between LP$_{01}$ and LP$_{11i}$ (circles), and between LP$_{01}$ and LP$_{11o}$ (crosses). The fiber is the same as in Fig. \ref{fig.coilmode}.}
\end{figure}

When a fiber is coiled all its modes are pushed toward the outside of the coil. One example is shown in Fig. \ref{fig.coilmode}.
Not surprisingly, the values of $F_M$ for the two orientations of LP$_{11}$ are shifted from the value in the same fiber without coiling, and the degeneracy between the two is broken. Figure \ref{fig.coilfreq} shows how $F_M$ shifts with a changing coil radius. These frequencies were computed using the assumption that the heat profile matches the irradiance profile, as was done in Fig. \ref{fig.freqvaeff}. They are meant as a qualitative illustration of the increase in $F_M$ with the stronger mode compression caused by tighter coiling.

\section{Dependence on pump wavelength}

If the pump light is detuned from the absorption peak at 976 nm, the reduced pump absorption coefficient will lead to lower upper state populations. This has the same effect as increasing the pump cladding diameter. It tends to increase hole burning and thus decrease $F_M$.

\section{Dependence on photo darkening}

The influence of photo darkening on $F_M$ depends strongly on the photo darkening model. If we use a model with signal absorption that is uniform across the core, the frequency should be pulled toward the value from Eq. (1), that is toward higher frequency than the full model. In addition, the added thermal lensing caused by absorptive heating will also pull toward higher frequency. However, if we use a photodarkening model in which the absorption is larger in areas with higher upper state population\cite{soderlund}, the absorption will be strongest near the outer edge of the core, and this will tend to reduce $F_M$. We have not tested the latter model, but we have verified the shift to higher frequencies with the former.

\section{Comparisons with measurements}

A major problem encountered in comparing our computed values of $F_M$ with measured frequencies is the general lack of information on the noise spectrum of pump or seed modulation. Both noise spectra are likely to fall in strength with increasing frequency. This means the observed frequency would tend to shift to lower values. Unfortunately, to date the reported frequency measurements have not included measurements of either the pump amplitude or spectral noise or the signal amplitude noise. It seems likely that pump or seed modulation is present, however, because both produce signature effects that seem to be present in measured performance. Thus there is circumstantial evidence of pump or seed modulation. We hope that in future experiments the fiber inputs will be carefully measured and documented. 

Although data is as yet not well enough documented for rigorous comparisons, we can test whether our predicted frequencies and trends are in reasonable agreement with measurements. We discuss published examples below.




%

\subsection{Example \#1}

Karow {\it et al.}\cite{karow} found $F_M\approx 1500$ Hz in an NKT fiber model DC-400-38-PZ-Yb. Properties of this fiber include $d_{\rm core}=d_{\rm dope}=38$ $\mu$m, $d_{\rm clad}=400$ $\mu$m, $NA=0.03$, $V=3.36$, $\lambda=1064$ nm, $d_{\rm mf}$=28 $\mu$m (extrapolated from DC-200-40 because 400-38 is not listed on NKT products), $L=1.45$ m, 9 W signal in, efficiency of conversion of absorbed pump of 0.80, a bend radius=20 cm, counter-pumped at 976 nm. They measured a signal noise spectrum that falls off with frequency roughly as $1/f$, but it has multiple prominent spikes, making meaningful comparison with our model questionable. The measured threshold is at an absorbed pump power of 94 W which produces 77 W of output signal. 

We compute $A_{\rm eff}$ of 990 $\mu$m$^2$for LP$_{01}$. The unusually low measured threshold power means there is insufficient heat load to cause significant thermal lensing. From the formula of Eq. (1) $F_M$ would be 1730 Hz. The lower measured frequency of $\approx 1500$ Hz is consistent with the frequency reduction expected from the population saturation effect. However, a more quantitative comparison requires a more thorough knowledge of the input signal and pump noise spectra.

\subsection{Example \#2}

Otto {\it et al.}\cite{otto} measured a frequency of 350 Hz for a counter-pumped NKT fiber model LPF45. Parameters of this fiber are $d_{\rm core}=81$ $\mu$m, $d_{\rm dope}=63$ $\mu$m, $d_{\rm clad}=255$ $\mu$m, $L=1.2$ m, $d_{\rm mf}=67.65$ $\mu$m, $A_{\rm eff}=3595$ $\mu$m$^2$ for LP$_{01}$. The measured threshold power was 200 W for the output signal. 

We modeled a counter-pumped LPF45 simulated by a step index with $V=10.2$ to match $A_{\rm eff}$ for LP$_{01}$, and found $F_M=490$ Hz. Other model inputs included: 10 W signal seed in LP$_{01}$ and $10^{-7}$ W frequency shifted light in LP11; 300 W pump input; 269 W signal output. 


A meaningful comparison between measurement and model is probably not possible in this case because the measured spectra are consistently highly structured, with three peaks at 300, 350, and 400  Hz, the most prominent being 350 Hz. We infer that the spectra do not reflect amplification of white noise but some other source of modulation, most likely either of the pump or the signal seed. The observed frequencies are not inconsistent with the modeled value of 490 Hz. A meaningful comparison will require careful measurements of the input pump and seed spectra.

\subsection{Example \#3}

Otto {\it et al.}\cite{otto} also measured the frequency for NKT fiber model DC-285/100 with $d_{\rm core}=100$ $\mu$m (step index), $d_{\rm dope}=100$ $\mu$m, $d_{\rm clad}=285$ $\mu$m, $d_{\rm mf}=76$ $\mu$m, $A_{\rm eff}=4536$ $\mu$m$^2$, $L=1.2$ m. The measured frequency was 220 Hz for a counter-pumped fiber at a threshold power of 100 W of signal output. The frequency according to Eq. (1) would be 375 Hz. 

We model a co-pumped, step index fiber with $NA=0.054$, signal seed of 10 W in LP$_{01}$, $10^{-5}$ W of frequency shifted LP$_{11}$, and 150 W of unmodulated pump. Our model predicts $F_M=325$ Hz for a co-pumped version which is substantially higher than the measured 220 Hz. A small amount of photo darkening might reduce the frequency to 220 Hz, or maybe the noise is 1/f like and so produces a lower frequency. Thermal lensing reduces $A_{\rm eff}$ by only 2\% at this power level, so that should not alter the frequency noticeably. Again, careful characterization of the input light is needed.

%

\subsection{Example \#4}

Ward {\it et al.}\cite{ward} measured a frequency of approximately 2000 Hz (with second, third, and fourth harmonics of 2000 Hz) for a fiber with $d_{\rm core}=39.5$ $\mu$m, $d_{\rm dope}=39.5$ $\mu$m, $d_{\rm mf}=30$ $\mu$m ($A_{\rm eff}=707$ $\mu$m$^2$), $d_{\rm clad}=329$ $\mu$m. The fiber was coiled with diameter of 53 cm. The signal seed was 30 W, the pump power was <1000 W at 976 nm, in a counter-pumped geometry. The measured threshold was at an output signal power of 380-500 W.

Eq. (1) suggests a gain peak at 2420 Hz, but hole burning would reduce this to nearer the measured 2000 Hz. The agreement between measurement and model is reasonable in this case. 

\section{Conclusions}

We showed using our detailed numerical model that the frequency of the highest mode coupling gain behaves in expected ways. If we ignore transverse hole burning so the oscillating portion of the heat profile is proportional to the product of fields in the two coupled modes, the predicted frequency scales as $1/A_{\rm eff}$ or $1/r_{\rm eff}^2$ as anticipated for a diffusive thermal wave. Furthermore, $F_M$ is quite close to the value given by Eq. (\ref{eq.89}) which is an estimate based on thermal diffusion across a heated cylinder of radius $r_{\rm eff}$.

We also showed that transverse hole burning tends to move the distribution of oscillating heat away from the axis of the core which tends to decrease $F_M$. Other effects tend to increase $F_M$. For example, confining the doping or reducing the mode size by thermal lensing both tend to increase the frequency of the gain peak. Coiling the fiber or cooling only one side both push the modes toward the outside of the bend and away from the cooled side, which tends to compress the modes and increase $F_M$.

Unfortunately, comparisons between measurement and model are not yet sufficiently advanced to verify these predicted effects. The measurements in most cases bear signs of input pump or signal modulation, and this means that the modulation spectra may strongly influence the measured oscillation frequencies. The obvious next step is more careful measurement of seed and pump modulation spectra.



\end{document}